\documentclass[prb, amsmath, amssymb, superscriptaddress, showpacs, reprint]{revtex4-1}

\usepackage{graphicx}				% Include figure files
\usepackage{dcolumn}

\begin{document}

\title{Pure dephasing in flux qubits due to flux noise with spectral density scaling as $1/ f^\alpha$}

\author{S.~M. Anton}
\affiliation{Department of Physics, University of California, Berkeley, CA 94720-7300, USA}

\author{C. M\"uller}
\affiliation{Institut f\"ur Theorie der Kondensierten Materie, Karlsruhe Institute of Technology, D-76128 Karlsruhe, Germany}
\affiliation{D\'{e}partement de Physique, Universit\'{e} de Sherbrooke, Sherbrooke, Qu\'{e}bec, Canada J1K 2R1}

\author{J.~S. Birenbaum}
\author{S.~R. O'Kelley}
\author{A.~D. Fefferman}
\affiliation{Department of Physics, University of California, Berkeley, CA 94720-7300, USA}
	
\author{D.~S. Golubev}
\affiliation{Institut f\"ur Nanotechnologie, Karlsruhe Institute of Technology, D-76021 Karlsruhe, Germany}
	
\author{G.~C. Hilton}
\author{H.-M. Cho}	
\author{K.~D. Irwin}
\affiliation{National Institute of Standards and Technology, Boulder, CO 80309-044, USA}
		
\author{F.~C. Wellstood}
\affiliation{Joint Quantum Institute, Department of Physics, University of Maryland, College Park, MD 20742, USA}

\author{Gerd~Sch\"on}
\affiliation{Institut f\"ur Nanotechnologie, Karlsruhe Institute of Technology, D-76021 Karlsruhe, Germany}
\affiliation{Institut f\"ur Theoretische Festk\"orperphysik, Karlsruhe Institute of Technology, D-76128 Karlsruhe, Germany}

\author{A. Shnirman}
\affiliation{Institut f\"ur Theorie der Kondensierten Materie, Karlsruhe Institute of Technology, D-76128 Karlsruhe, Germany}
	
\author{John Clarke}
\affiliation{Department of Physics, University of California, Berkeley, CA 94720-7300, USA}

\date{\today}

\begin{abstract}
For many types of superconducting qubits, magnetic flux noise is a source of pure dephasing. Measurements on a representative dc superconducting quantum interference device (SQUID) over a range of temperatures show that $S_\Phi(f) = A^2/(f/1~\hbox{Hz})^\alpha$, where $S_\Phi$ is the flux noise spectral density, $A$ is of the order of 1~$\mu\Phi_0 \, \hbox{Hz}^{-1/2}$ and $0.61 \leq \alpha \leq 0.95$; $\Phi_{0}$ is the flux quantum. For a qubit with an energy level splitting linearly coupled to the applied flux, calculations of the dependence of the pure dephasing time $\tau_\phi$ of Ramsey and echo pulse sequences on $\alpha$ for fixed $A$ show that $\tau_\phi$ decreases rapidly as $\alpha$ is reduced. We find that $\tau_\phi$ is relatively insensitive to the noise bandwidth, $f_1 \leq f \leq f_2$, for all $\alpha$ provided the ultraviolet cutoff frequency $f_2 > 1/\tau_\phi$. We calculate the ratio $\tau_{\phi,E} / \tau_{\phi,R}$ of the echo ($E$) and Ramsey ($R$) sequences, and the dependence of the decay function on $\alpha$ and $f_2$. We investigate the case in which $S_\Phi(f_0)$ is fixed at the ``pivot frequency'' $f_0 \neq 1$~Hz while $\alpha$ is varied, and find that the choice of $f_0$ can greatly influence the sensitivity of $\tau_{\phi,E}$ and $\tau_{\phi,R}$ to the value of $\alpha$. Finally, we present calculated values of $\tau_\phi$ in a qubit corresponding to the values of $A$ and $\alpha$ measured in our SQUID.
\end{abstract}

%\keywords{}

\maketitle

\section{Introduction}

The dynamics of superconducting quantum bits (qubits)~\cite{Clarke:N:2008}---broadly classified as charge qubits~\cite{Nakamura:PRL:1997}, flux qubits~\cite{vanderWaal:S:2000} and phase qubits~\cite{Martinis:PRL:2002}---can be characterized by two times: the relaxation time $T_1$ and the pure dephasing time $\tau_\phi$~\cite{Makhlin:RMP:2001}. The time $T_1$ required for a qubit to relax from its first excited state to its ground state is determined by the strength of environmental fluctuations at a frequency corresponding to the energy level splitting $\nu_{01}$ of the two states. The decoherence time $T_2$, over which the phase of superpositions of two eigenstates becomes randomized, has two contributions: $1/T_2 = 1/(2 \, T_1) + 1/\tau_\phi$. The pure dephasing time $\tau_\phi$ is limited by fluctuations in $\nu_{01}$, due predominantly to fluctuations in magnetic flux in the case of flux qubits.

Excess low frequency flux noise was first identified in dc Superconducting QUantum Interference Devices (SQUIDs)~\cite{Koch:JLTP:1983}. Measurements at millikelvin temperatures~\cite{Wellstood:APL:1987, Drung:IEEE:2011, Sendelbach:PRL:2008, Sank:arXiv:2011} reveal a power spectrum $S_\Phi(f)$ scaling as $1/f^\alpha$ ($f$ is frequency), with an amplitude at 1~Hz typically of the order of $1~\mu\Phi_0 \, \text{Hz}^{-1/2}$, that is surprisingly uniform for SQUID washers of widely differing geometries that are fabricated with a variety of materials. Here, $\Phi_0 \equiv h/2e$ is the flux quantum.

Flux noise is believed to arise from the random reversal of electron spins at the interface between a superconducting film and an insulator~\cite{Koch:PRL:2007, Faoro:PRL:2008, Choi:PRL:2009}. The areal density of independent spins required to account for the observed flux noise is about $5\times 10^{17} ~ \text{m}^{-2}$, a value that has been corroborated by observations of paramagnetism in SQUIDs~\cite{Sendelbach:PRL:2008} and normal metal rings~\cite{Bluhm:PRL:2009}. Recent experiments on anti-correlations of flux noise have confirmed the surface spin model~\cite{Yoshihara:PRB:2010, Gustavsson:2011}. An unambiguous understanding of the mechanism by which the spins produce $1/f$ flux noise, however, has yet to be developed. Recent proposals include spin clusters~\cite{Sendelbach:PRL:2009}, spin glasses~\cite{Chen:PRL:2010}, fractal spin clusters~\cite{Kechedzhi:2011}, and hyperfine interactions\cite{Jiansheng:arXiv:2011}; the models in Refs.~\cite{Chen:PRL:2010, Kechedzhi:2011} suggest that $\alpha$ may differ from unity.

Measurements of $\tau_{\varphi}$ in flux qubits~\cite{Yoshihara:PRB:2010, Gustavsson:2011, Yoshihara:PRL:2006, Kakuyanagi:PRL:2007, Lanting:PRB:2009} and phase qubits~\cite{Bialczak:PRL:2007} have been used to infer the magnitude of the flux noise in these devices, under the assumption that the spectral density of the flux noise scaled as $1/f$. In this paper we first present measurements of flux noise spectral densities scaling as $1/f^\alpha$ in which the exponent $\alpha$ deviates markedly from unity. We then show theoretically that such deviations strongly impact $\tau_\varphi$: the value of $\tau_\varphi$ decreases markedly with decreasing $\alpha$. Additionally, we examine the influence of $\alpha$ and noise bandwidth on $\tau_\phi$, the ratio $\tau_{\phi,E} / \tau_{\phi,R}$ obtained in echo ($E$) and Ramsey ($R$) pulse sequences, and the functional dependence of the decay function. Finally, we calculate the predicted $\tau_\phi$ for values of $A$ and $\alpha$ obtained in our measurements.

\section{Experimental procedures and results}

\begin{figure}[b]
\begin{center}
\includegraphics[trim = 0 0.2in 0 0]{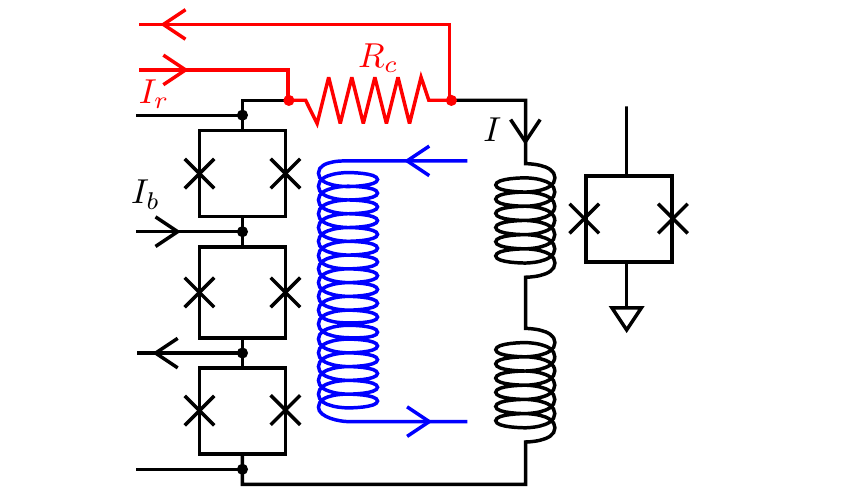}
\caption{(Color online) Circuit schematic of measurement system for three measured SQUIDs. Measurement of the middle SQUID is shown (see text). A single coil applies flux to all SQUIDs.}
\label{fig:schematic}
\end{center}
\end{figure}

We measured the flux noise spectral densities of Nb-based dc SQUIDs, fabricated using a 50~ A/cm$^2$ Nb/AlO$_\text{x}$/Nb trilayer process~\cite{Sauvageau:1995}. Each $2.5~\times~2.5~\mu\text{m}^2$ junction was shunted with a $2.5~\Omega$ PdAu resistor to eliminate hysteresis on the current-voltage ($I$-$V$) characteristic. Up to six SQUIDs, connected in series, were in turn connected in series with a compensating resistor $R_{c} \approx 0.45~\Omega$ and the superconducting input coil of a readout SQUID, operated in a flux-locked loop (Fig.~\ref{fig:schematic}). To measure the noise in a given SQUID, we applied a current $I_b$ sufficient to produce a voltage $V$ of typically $5~\mu \mbox{V}$ across it. The resulting static current around the circuit was cancelled by an appropriate current $I_r$ in $R_c$ to ensure that the remaining SQUIDs remained in the zero-voltage state. In addition, a choke inductor was used to decouple oscillations at the Josephson frequency between the measured and readout SQUIDs. Since $R_c$ was much less than the dynamic resistance of any given SQUID, the SQUID was effectively voltage biased. Fluctuations in the critical current of the measured SQUID induced a current noise $I$ with spectral density $S_{I}(f)$ in the input coil of the readout SQUID. We inferred the flux noise from $S_{\Phi}(f) = S_{I}(f) / [ (\partial I / \partial \Phi)_{V} ]^{2}$, where $\Phi$ is the flux applied via an external coil to the measured SQUID and $(\partial I / \partial \Phi)_{V}$ was determined separately. We also measured the critical current noise of the junctions\cite{Koch:JLTP:1983} in each SQUID biased at $n\Phi_0$ ($n$ is an integer) so that $(\partial I / \partial \Phi)_V = 0$. This noise was negligible compared with the current noise produced by the flux noise for large values of $(\partial I/\partial \Phi)_V$. The experiment was mounted in a lead-coated copper box surrounded by a cylindrical lead shield inside a cryoperm shield, and cooled with a dilution refrigerator. All leads were heavily filtered.

\begin{figure}[t]
\begin{center}
\includegraphics[trim = 0 0.2in 0 0]{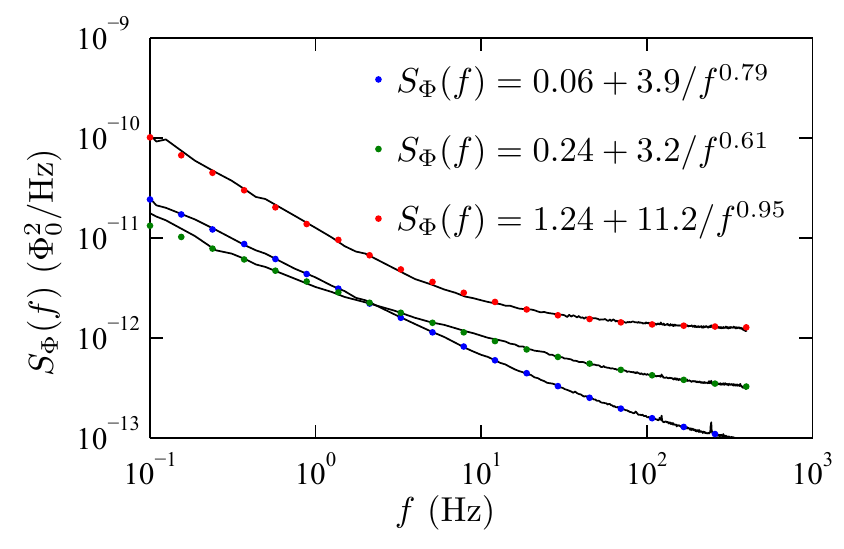}
\caption{(Color online) Measured and fitted flux noise spectra. Measured spectra at 0.2, 1.3, and 4.5~K are continuous curves; dots are fits to Eq.~\eqref{eq:SPhi}. Quoted values in the fits are in units of $(\mu\Phi_0)^2/\text{Hz}$.}
\label{fig:NoiseSpectra}
\end{center}
\end{figure}

Figure~\ref{fig:NoiseSpectra} shows power spectra of a single SQUID at three different temperatures. The inner and outer dimensions of the washer were 50 and $90~\mu\text{m}$, respectively. At higher frequencies, the spectra begin to flatten out due to white current noise from the SQUID shunt resistors, which dominates that from $R_{c}$.
The spectra were fitted to the form
\begin{equation}
	S(f) = A^{2} / (f / 1~\text{Hz})^\alpha + C^2
	\label{eq:SPhi}
\end{equation}
with parameters $A$ and $C$ for the amplitude of the ``$1/f$'' flux noise and white noise, respectively. We found that the exponent $\alpha$ of the $1/f^\alpha$ noise can be far from unity, varying from $0.61$ to $0.95$. We remark that these values are representative of measurements on about 20~SQUIDs. Since the junctions in flux qubits are not resistively shunted, we shall focus on dephasing from the term $A^2/(f / 1~\text{Hz})^\alpha$.

\section{Theoretical results}

\subsection{Model}

Since $\alpha$ can evidently be much less than unity, it is natural to ask what impact this has on the pure dephasing of flux qubits. Low frequency flux noise modulates the energy splitting of the ground and first excited states of a flux qubit, $h \nu_{01} = [\Delta^2 + \epsilon^2(\Phi)]^{1/2}$, via the bias energy $\epsilon(\Phi)$. The bias energy is the energy difference between the two states with persistent currents $\pm I_q$ when there is no tunneling between them ($\Delta=0$)~\cite{Mooij:PRB1999}.  Here, $\epsilon = 2I_q (\Phi - \Phi_0/2)$ or equivalently, $I_q \equiv \tfrac{1}{2}(\partial \epsilon/ \partial \Phi)$.

We define the sensitivity of the splitting to a change in $\Phi$ in terms of the longitudinal sensitivity of the qubit to flux noise,
\begin{equation}
D_\Phi \equiv \partial \nu_{01}/\partial \Phi = (1/h) (\partial \epsilon / \partial \Phi) \epsilon/(\Delta^2 + \epsilon^2)^{1/2}\ .
\label{eq:DPhi}	
\end{equation}
To first order, there is no dephasing from flux noise at the degeneracy point, $\Phi = \Phi_0/2$, where $\epsilon$ vanishes. In this paper, however, we consider the limit $\epsilon/\Delta \gg 1$, far from the degeneracy point, at which $\partial \nu_{01} / \partial \Phi = (1/h)\partial \epsilon / \partial \Phi = 2I_q/h$. We assume that, in this limit, $1/\tau_\phi \gg 1/2T_1$ so that the measured dephasing time arises only from pure dephasing. We adopt the value $D_\Phi = 10^{12}~\text{Hz}/\Phi_0$, corresponding to the typical value~\cite{Yoshihara:PRL:2006, Kakuyanagi:PRL:2007} $I_q \approx 0.3~\mu\text{A}$. Furthermore, based on the empirical observation that $S_\Phi(1~\text{Hz})$ is relatively constant among a wide variety of SQUIDs, we assume that $A = 1~\mu\Phi_0 \, \text{Hz}^{-1/2}$ regardless of the value of $\alpha$. We consider noise fixed at a frequency other than 1~Hz in Sec.~\ref{sec:pivot}.

The modulation of $\nu_{01}$ by flux noise leads to an accumulation of phase error and thus to dephasing. The rate at which the dephasing occurs varies between different types of pulse sequences. For example, in a Ramsey sequence~\cite{Ramsey:PR:1950} the qubit is excited by a microwave $\pi/2$ pulse from the ground state into a superposition of ground and excited states. After a time $t$ another $\pi/2$ pulse is applied and the qubit state is measured. The results of many measurements with fixed $t$ are averaged and $t$ is varied from $t \ll \tau_\phi$ to $t \gg \tau_\phi$ to obtain the decay function $g(t)$. Here, we define the dephasing time as $g(\tau_\phi) \equiv 1/e$. To eliminate dephasing due to flux fluctuations between pulse sequences, one implements an echo sequence in which a $\pi$ pulse is inserted midway between the two $\pi/2$ pulses~\cite{Hahn:PR:1950}. In general, the echo sequence yields a dephasing time greater than that of the Ramsey: $\tau_{\phi,E} > \tau_{\phi,R}$.

The sensitivity of the Ramsey and echo sequences to noise are described by the weighting function $W(f,t)$ given by~\cite{Martinis:PRL:2002, Ithier:PRB:2005}
\begin{align}
W_R(f,t) = \frac{\sin^2(\pi f t)}{(\pi f t)^2} \, , && W_E(f,t) = \frac{\sin^4(\pi f t/2)}{(\pi f t /2)^2}.
\end{align}
For the Ramsey sequence with $f \ll 1/t$, we see that $W_R \approx 1$, whereas for $f \gg 1/t$, $W_R$ falls as $1/f^2$. Consequently, we expect the dominant contributions to the Ramsey dephasing time to arise from noise at frequencies $f \lesssim 1/\tau_{\phi,R}$. In contrast, for the echo sequence $W_E$ scales as $f^2$ for $f \ll 1/t$ and as $1/f^2$ for $f \gg 1/t$. In this case, we expect the dominant contribution to the dephasing time to be from noise at frequencies $f \approx 1/\tau_{\phi,E}$.

The decay function $g(t)$ is calculated by ensemble averaging over the entire measurement time, yielding~\cite{Martinis:PRL:2002, Ithier:PRB:2005}
\begin{equation} \label{eq:decay_function}
g(t) = \exp\left[ -t^2 (2\pi D_\Phi)^2 \int_{f_1}^{f_2} df S_{\Phi}(f) W(f,t) \right] \, .
\end{equation}
Here, the symmetrized noise power is defined as $S_{\Phi}(f) \equiv (1/2) \int dt  \left\{\langle \Phi(t) \Phi(0)\rangle +\langle \Phi(0) \Phi(t)\rangle \right\} e^{-2\pi i f t}$, which we replace with the observed spectrum: $S_\Phi(f) = A^2/(f/(1~\text{Hz}))^\alpha$; $f_1$ and $f_2$ are cutoff frequencies limiting the noise frequency bandwidth to which the qubit is sensitive. Independent of the particular pulse sequence, the infrared cutoff $f_1$ is set by the entire measurement time $T$ taken to acquire sufficient statistics to determine the decay function $g(t)$, that is $f_1 = 1/T$, where $T$ may range from, say, 1~ms to 1000~s. What determines the ultraviolet cutoff $f_2$, however, is less clear. Recent experiments~\cite{Bylander:2011:NP, Slichter:arXiv:2012} indicate that flux noise can not only extend to very high frequencies (in one case in excess of 1~GHz), but maintain its nonunity value of $\alpha$ out to $f_2$.

% Figure 3:

\begin{figure}[b]
\begin{center}
\includegraphics[trim = 0 0.2in 0 0]{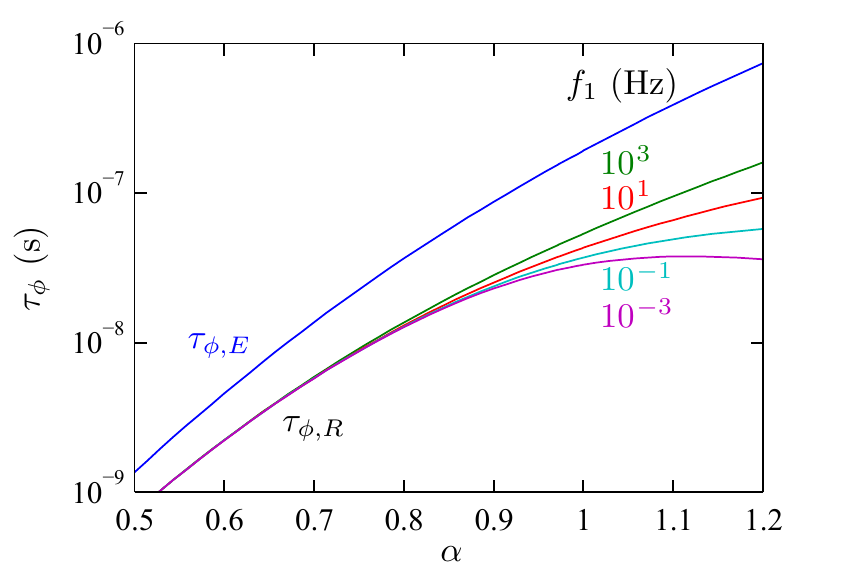}
\caption{(Color online) Computed values of $\tau_{\phi,R}$ and $\tau_{\phi,E}$ vs $\alpha$ for $f_1 = 10^{-3}$, 0.1, 10, and $10^3$~Hz and $f_2 \rightarrow \infty$.}
\label{fig:tauphi_vs_alpha}
\end{center}
\end{figure}

\subsection{Dephasing times versus $\alpha$}

As is evident from Eq.~\eqref{eq:decay_function}, a nonunity value of $\alpha$ will affect the integral in a complicated way. Figure~\ref{fig:tauphi_vs_alpha} shows computed dephasing times for both sequences versus $\alpha$ for $f_2 \rightarrow \infty$ and $f_1 = 10^{-3}$, $10^{-1}$, $10^1$, and $10^{3}~\text{Hz}$. The effect of changing $\alpha$ is substantial: both $\tau_{\phi,R}$ and $\tau_{\phi,E}$ increase by an order of magnitude as $\alpha$ is varied from 0.6 to 0.9. By comparison, we find that an order of magnitude change in $A$ for a given value of $\alpha$ also changes $\tau_\phi$ by an order of magnitude. Figure~\ref{fig:tauphi_vs_alpha} further shows that, because of its insensitivity to low frequency noise, the echo sequence yields significantly longer dephasing times for all $\alpha$. Finally, while $\tau_{\phi,E}$ is insensitive to changes in $f_1$ for $f_1 \ll 1/\tau_{\phi,E}$ (equivalently $T \gg \tau_{\phi,E}$), $\tau_{\phi,R}$ becomes increasingly sensitive as $\alpha$ increases.

% Figure 4:

\begin{figure}[t]
\begin{center}
\includegraphics[trim = 0 0.2in 0 0]{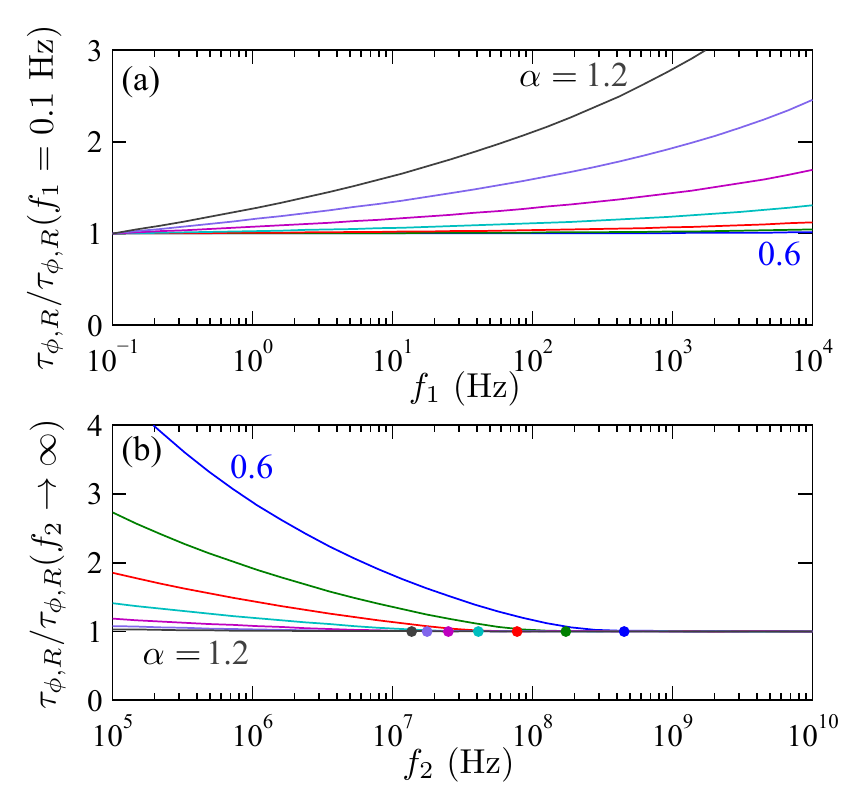}
\caption{(Color online) Normalized Ramsey dephasing times for $0.6 \le \alpha \le 1.2$ in steps of 0.1. (a) $\tau_{\phi,R}(f_1) / \tau_{\phi,R}(f_1 = 0.1~\text{Hz})$ versus $f_1$ for $f_2 \rightarrow \infty$ and (b) $\tau_{\phi,R}(f_2) / \tau_{\phi,R}(f_2 \rightarrow \infty)$ versus $f_2$. The colored dots in (b) are placed at $f_2 = 1/\tau_{\phi,R}(f_2 \rightarrow \infty)$, above which $\tau_{\phi,R}$ displays no dependence on $f_2$ (see text).}
\label{fig:tauphiR_vs_f}
\end{center}
\end{figure}

\subsection{Dephasing times versus cutoff frequencies}

We now examine more quantitatively the sensitivity of $\tau_\phi$ to changes in both $f_1$ and $f_2$ for various values of $\alpha$. For the Ramsey sequence with $f_2 \rightarrow \infty$, Fig.~\ref{fig:tauphiR_vs_f}(a) shows $\tau_{\phi,R}$, normalized to $\tau_{\phi,R}(f_1 = 0.1~\text{Hz})$, versus $f_1$ for $0.6 \le \alpha \le 1.2$. We again see that the sensitivity of $\tau_{\phi,R}$ to $f_1$ increases with increasing $\alpha$. Even so, for $\alpha = 1.2$, $\tau_{\phi,R}$ changes by a factor of only 4 when $f_1$ is varied from $0.1$ to $10^4$~Hz.

To explore the effect of $f_2$ on $\tau_{\phi,R}$, we fix $f_1 = 1~\text{Hz}$ and vary $f_2$, plotting $\tau_{\phi,R}(f_2) / \tau_{\phi,R}(f_2 \rightarrow \infty)$ for $0.6 \le \alpha \le 1.2$ [Fig.~\ref{fig:tauphiR_vs_f}(b)]. We see that the sensitivity of $\tau_{\phi,R}$ to $f_2$ increases for decreasing $\alpha$. Furthermore, Fig.~\ref{fig:tauphiR_vs_f}(b) shows that $\tau_{\phi,R}$ is insensitive to the particular value of $f_2$ for $f_2 \gg 1/\tau_{\phi,R}(f_2 \rightarrow \infty)$, simply because the Ramsey sequence is insensitive to noise for $f \gg 1/\tau_{\phi,R}$. However, as $f_2$ decreases through $1/\tau_{\phi,R}(f_2 \rightarrow \infty)$ a non-negligible amount of noise to which the qubit is sensitive is effectively eliminated, thereby reducing the total integrated noise and increasing $\tau_{\phi,R}(f_2)$. This effect is greater for small $\alpha$, where $S_\Phi$ decreases with $f$ more slowly and contributes to dephasing out to a higher frequency.

\begin{figure}[t]
\begin{center}
\includegraphics[trim = 0 0.2in 0 0]{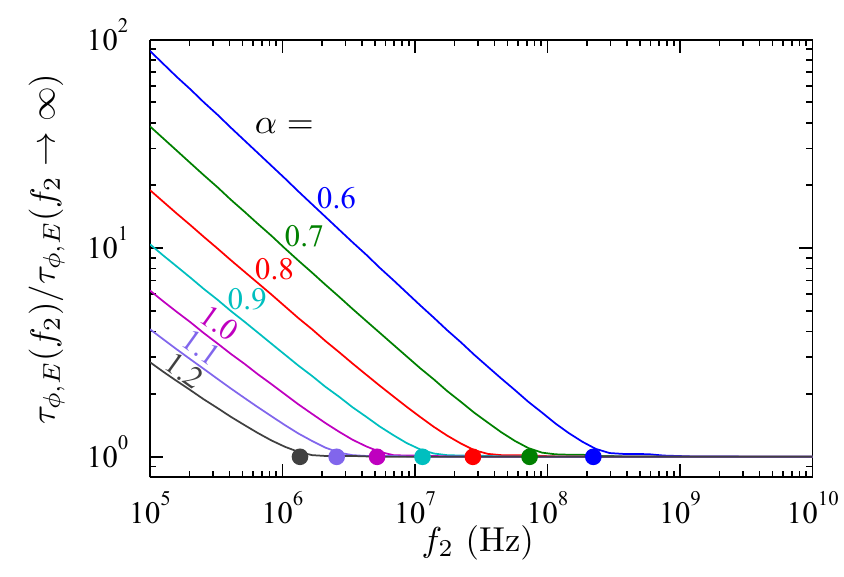}
\caption{(Color online) Computed values of $\tau_{\phi,E}(f_2) / \tau_{\phi,E}(f_2 \rightarrow \infty)$ vs $f_2$. Lower cutoff frequency $f_1 = 1~\text{Hz}$ and $0.6 \le \alpha \le 1.2$ in steps of 0.1. The colored dots are placed at $f_2 = 1/\tau_{\phi,E}(f_2 \rightarrow \infty)$, above which $\tau_{\phi,E}$ displays no dependence on $f_2$ (see text).}
\label{fig:tauphiE_vs_f2}
\end{center}
\end{figure}

We perform a similar analysis of the sensitivity of $\tau_{\phi,E}$ to the value of $f_2$. In Fig.~\ref{fig:tauphiE_vs_f2} we plot $\tau_{\phi,E}(f_2) / \tau_{\phi,E}(f_2 \rightarrow \infty)$ versus $f_2$ for $f_1 = 1$~Hz. As with the Ramsey sequence, we find that $\tau_{\phi,E}$ is insensitive to $f_2$ for $f_2 \gg 1/\tau_{\phi,E}(f_2 \rightarrow \infty)$. Indeed, since $\tau_{\phi,E}$ is dominated by noise at $f \approx 1/\tau_{\phi,E}$, this result as we expect. Also in analogy with the Ramsey sequence, $\tau_{\phi,E}$ is more sensitive to $f_2$ for small $\alpha$. Unlike the Ramsey sequence, however, where the dephasing is sensitive to frequencies over a large bandwidth ($f_1$ to $1/\tau_{\phi,R}$), the echo sequence is sensitive to noise only in a narrow bandwidth around $1/\tau_{\phi,E}$, making $\tau_{\phi,E}$ much more sensitive to changes in $f_2$ for $f_2 \approx 1/\tau_{\phi,E}$. Here, $\tau_{\phi,E}$ increases by an order of magnitude for a two-order-of-magnitude decrease in $f_2$.

% Figure 5

% Figure 6

\begin{figure}[b]
\begin{center}
\includegraphics[trim = 0 0.2in 0 0]{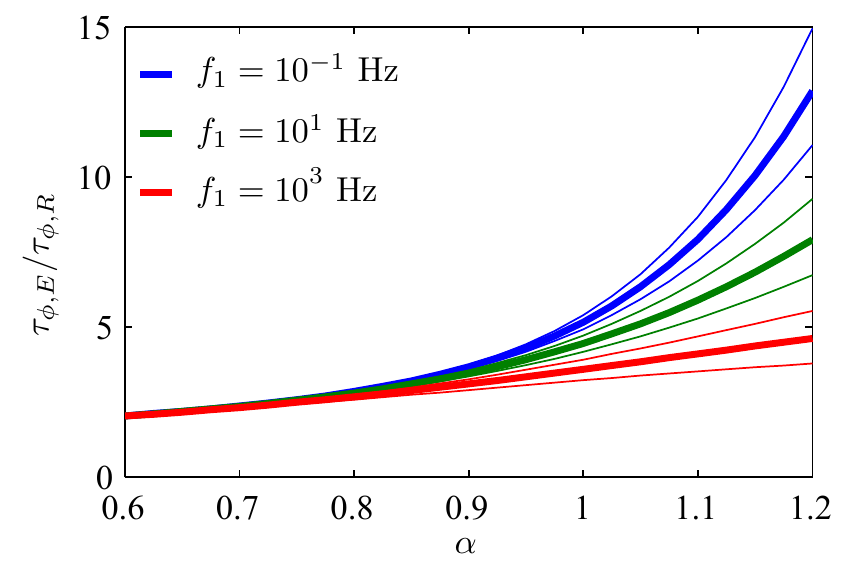}
\caption{(Color online) Ratio $\tau_{\phi,E} / \tau_{\phi,R}$ vs $\alpha$. Lower cutoff frequency $f_1 = 10^{-1}$, $10^1$, and $10^3$~Hz and $f_2 \rightarrow \infty$. The thin upper, heavy middle, and thin lower lines correspond to $A D_\phi / (10^6~\text{Hz}^{1/2}) = 0.2$, $1$, and $5$, respectively.}
\label{Fig:ratio}
\end{center}
\end{figure}

\subsection{The ratio $\tau_{\phi,E} / \tau_{\phi,R}$}

Since the value $D_\phi$ can vary significantly between flux qubits, we consider the ratio $\tau_{\phi,E} / \tau_{\phi,R}$, which has the advantage of being rather insensitive to the precise values of both $A$ and $D_\phi$. We compute these times using Eq.~(4), which shows that the decay function $g(t)$ depends only on the product $A D_\phi$. To explore the dependence of the ratio on $\alpha$, we compute $\tau_{\phi,E} / \tau_{\phi,R}$ versus $\alpha$ for $f_2 \rightarrow \infty$ (equivalent to $f_2 \gg 1/\tau_\phi$) and $f_1 = 10^{-1}$, $10^1$, and $10^{3}~\text{Hz}$. Furthermore, for each value of $f_1$ we perform the calculation for $A D_\phi / (10^6~\text{Hz}^{1/2}) = 0.2$, $1$, and $5$. The results are shown in Fig.~\ref{Fig:ratio}.

We first examine the dependence on $\alpha$. As $\alpha$ increases, noise at frequencies much greater than 1~Hz falls quickly, so that $\tau_{\phi,E}$ increases rapidly. Conversely, noise at low frequencies near 1~Hz changes little as $\alpha$ changes. The Ramsey dephasing time is sensitive to a large noise bandwidth where a significant contribution comes from frequencies near $f_1$. Therefore, as $\alpha$ increases we expect $\tau_{\phi,R}$ to increase less rapidly than $\tau_{\phi,E}$, explaining the increasing trend of $\tau_{\phi,E} / \tau_{\phi,R}$

For small, fixed values of $\alpha$, changing the value of $f_1$ changes the ratio only slowly because both $\tau_{\phi,E}$ and $\tau_{\phi,R}$ are limited by noise at $f \gg f_1$. As the value of $\alpha$ increases, however, an increasing contribution to dephasing in the Ramsey sequence arises from lower frequencies $f \approx f_1$. Therefore, for large, fixed values of $\alpha$, increasing $f_1$ has the effect of removing a significant noise contribution, thereby increasing $\tau_{\phi,R}$ and decreasing the ratio $\tau_{\phi,E} / \tau_{\phi,R}$. We remark that since $f_1$ is an experimentally variable parameter, measuring $\tau_{\phi,E} / \tau_{\phi,R}$ for several different measurement times may shed light on the value of $\alpha$.

Finally, we see that the $\tau_{\phi,E} / \tau_{\phi,R}$ is moderately sensitive to the product $AD_\phi$ only for $\alpha \gtrsim 1$. However, additional calculations show that, for $f_2 \lesssim 1/\tau_\phi$, the ratio becomes extremely sensitive to the particular value of $A D_\phi$.

\subsection{Dependence of decay function on $\alpha$ and ultraviolet cutoff frequency}

The decay function is of particular interest experimentally, since it can be measured directly. In general, the decay function of $T_1$-limited processes is a simple exponential, that is $g(t) = \exp(-t/T_1)$. However, the decay function of pure dephasing processes is more complicated and can be characterized as $g(t) \equiv \exp(-\chi(t))$, where $\chi(t)$ can contain terms that are higher order in $t$.

% Figure 7:

\begin{figure}[b]
\begin{center}
\includegraphics[trim = 0 0.2in 0 0]{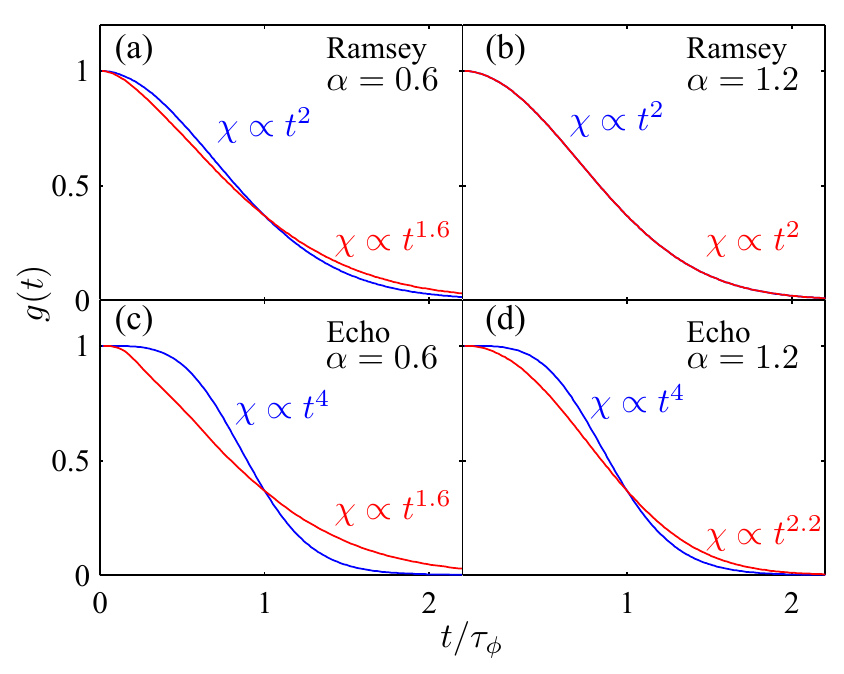}
\caption{(Color online) Computed decay function $g(t)$ versus $t/\tau_\phi$ for (a) and (b) Ramsey sequences and (c) and (d) echo sequences with $\alpha = 0.6$ and 1.2. In the red trace, $f_2 \gg 1/\tau_\phi(f_2 \rightarrow \infty)$; in the blue trace $f_2 \ll 1/\tau_\phi(f_2 \rightarrow \infty)$.}
\label{decay_function}
\end{center}
\end{figure}

Here, we examine the functional dependence of $\chi(t)$ for both pulse sequences. In each case, we find that $\chi(t) \propto t^\gamma$, where $\gamma$ can take two values ($\gamma_1$ and $\gamma_2$) within a single sequence, separated by a characteristic time set by $1/f_2$: $\chi(t \ll 1/f_2) \propto t^{\gamma_1}$ and $\chi(t \gg 1/f_2) \propto t^{\gamma_2}$. For the Ramsey sequence, $\gamma_1 = 2$ and $\gamma_2 = 1+\alpha$ for $\alpha \le 1$ and $\gamma_2 = 2$ for $\alpha > 1$. For the echo sequence, $\gamma_1 = 4$ and $\gamma_2 = 1+\alpha$. These results reveal two experimentally relevant insights. First, for $t \gg 1/f_2$, $\gamma_2$ depends on $\alpha$. Thus, if $\tau_\phi \gg 1/f_2$, a careful fit of the experimentally observed decay envelope may shed light on the value of $\alpha$. Second, the functional form of $g(t)$ can reveal information about $f_2$. For example, if one does not observe that $\chi(t) \propto t^4$ in an echo experiment, $f_2$ must be as high as $1/\tau_{\phi,E}$, establishing an important lower bound on the bandwidth of the flux noise.

Figure~\ref{decay_function} emphasizes the above statements, showing $g(t)$ plotted for both sequences for $\alpha = 0.6$ and 1.2, and for $f_2$ both above and below $1/\tau_\phi(f_2 \rightarrow \infty)$, thereby showing both $\gamma_1$ and $\gamma_2$ dependence. In Figs.~\ref{decay_function}(a) and \ref{decay_function}(b) we plot $g(t)$ for the Ramsey sequence with $\alpha = 0.6$ and 1.2. We note that difference between the functional dependencies of the two traces in Fig.~\ref{decay_function}(a) is slight, and would be nearly impossible to measure experimentally. In Fig.~\ref{decay_function}(b) there is no functional difference since $\gamma_1 = \gamma_2$. Figures~\ref{decay_function}(c) and \ref{decay_function}(d) show $g(t)$ for the echo sequence for $\alpha = 0.6$ and 1.2. The difference is more dramatic since $\gamma_1 = 4$ is so large. In this case, such a difference might be experimentally observable.

% Figure 8:

\begin{figure}[b]
\begin{center}
\includegraphics[trim = 0 0.2in 0 0]{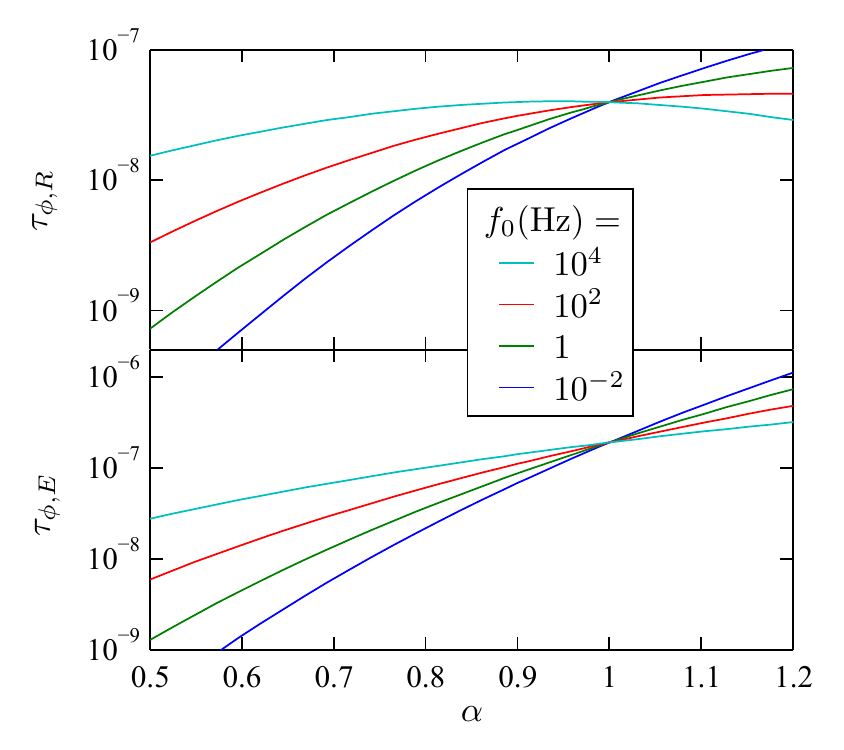}
\caption{(Color online) Computed dephasing times $\tau_\phi$ vs $\alpha$ for $f_1 = 1~$Hz and $f_2\rightarrow\infty$. (a) Ramsey and (b) echo pulse sequences for fixed $S_\Phi(f_0)$, where the pivot frequency $f_0 = 10^{-2}, 1, 10^2,$ and $10^4$~Hz.}
\label{Fig:pivot}
\end{center}
\end{figure}

\subsection{\label{sec:pivot}$S_\Phi(f)$ pivoting about $f_0\neq1$~Hz as $\alpha$ is varied}

As mentioned previously, there is no \emph{a priori} reason to hold $S_\Phi(1~\text{Hz})$ fixed as $\alpha$ is varied; the choice is based on the empirical observation that values of $S_\Phi(1~\text{Hz})$ are relatively uniform across a wide variety of devices and measured $\alpha$. To explore the sensitivity of our calculations to this assumption, we calculated the dephasing times for both sequences versus $\alpha$ for fixed $S_\Phi(f_0)$, where $f_0 = 10^{-2}, 1, 10^2,$ and $10^4$~Hz. Conceptually, the spectra can be imagined as pivoting as $\alpha$ changes about a fixed spectral density $S_\Phi(f_0)$ at frequency $f_0$. In order to normalize the magnitude of each set of curves corresponding to a particular $f_0$, we choose as a convention that $S_\Phi(1~\text{Hz}) = A^2$ when $\alpha = 1$, regardless of the value of $f_0$, that is $S_\Phi(f) = (A^2/f_0)(f/f_0)^{-\alpha}$. This convention is based loosely on empirical observation; it does not significantly change the dependence of $\tau_\phi$ on $\alpha$, but merely sets the absolute scale.

The results of these calculations, plotted in Fig.~\ref{Fig:pivot}, show a dramatic effect, both qualitatively and quantitatively, on the dependence of $\tau_{\phi,R}$ and $\tau_{\phi,E}$ on $\alpha$. For the Ramsey sequence [Fig.~\ref{Fig:pivot}(a)], the general trend of increasing $\tau_{\phi,R}$ is significantly altered as $f_0$ increases and even becomes nonmonotonic for $f_0 = 10^4$~Hz. In addition, for small values of $\alpha$, $\tau_{\phi,R}$ increases dramatically as $f_0$ increases. We note that, because of our normalization condition, the curves intersect at $\alpha = 1$. Calculations for the echo sequence are shown in Fig.~\ref{Fig:pivot}(b), which shows a similar dependence of $\tau_{\phi,E}$ on $f_0$. For both sequences, the dependence of $\tau_\phi$ on $\alpha$ is minimal for $f_0 = 10^4$~Hz, the highest computed $f_0$. This dependence is easily understood for the echo sequence, which is sensitive only to noise at $f \approx 1/\tau_{\phi,E}$. As $f_0$ approaches $1/\tau_{\phi,E}$, the effect of $\alpha$ eventually becomes negligible. In fact, if $f_0$ were to exceed $1/\tau_{\phi,E}$, the trend in $\alpha$ would actually reverse. The Ramsey sequence, however, is sensitive to a larger noise bandwidth and has a correspondingly more complicated dependence, exhibited by its nonmonotonic behavior for large values of $f_0$.

\subsection{Tabulated dephasing times}

Finally, we use our theoretical prediction of a strong dependence of the dephasing times on $\alpha$ to calculate $\tau_{\phi,R}$ and $\tau_{\phi,E}$ for the experimental values of $A$ and $\alpha$ shown in Fig.~\ref{fig:NoiseSpectra}. We assume $f_1 = 1$~Hz and $f_2 \rightarrow \infty$. The results are shown in the upper section of Table~\ref{tab:computed_tau_phi}. We see that, despite having the largest value of the flux noise magnitude $A$, the spectrum with the highest value of $\alpha$, 0.95, yields the longest dephasing times. This result emphasizes a crucial point: simply lowering the flux noise magnitude $A$ while keeping $\alpha$ constant may not be the most effective avenue to increasing $\tau_\phi$. The middle section of Table~\ref{tab:computed_tau_phi} shows the effect on $\tau_{\phi,R}$ and $\tau_{\phi,E}$ of a ten-fold reduction in $A$ for fixed $\alpha$. The factors by which $\tau_{\phi,R}$ and $\tau_{\phi,E}$ increase are comparable and decrease as $\alpha$ increases, from about 17 ($\alpha = 0.61$) to about 11 ($\alpha = 0.95$). The values of $\tau_{\phi,R}$ and $\tau_{\phi,E}$ for $A = 1~\mu\Phi_0 \, \text{Hz}^{-1/2}$ are shown in the lower section of Table~\ref{tab:computed_tau_phi}. As expected, $\tau_{\phi,R}$ and $\tau_{\phi,E}$ increase dramatically as $\alpha$ increases from 0.61 to 0.95.

% Table I

\begin{table}[b]
\caption{\label{tab:computed_tau_phi}Computed $\tau_{\phi,R}$ and $\tau_{\phi,E}$ with $f_1 = 1$~Hz and $f_2 \rightarrow \infty$ for flux qubits with $I_q = 0.3~\mu\text{A}$. Upper section: values of $A$ and $\alpha$ from Fig.~\ref{fig:NoiseSpectra}; middle section: $A$ reduced by factor of 10, $\alpha$ unchanged; lower section: $A$ set equal to $1~\mu\Phi_0 \, \text{Hz}^{-1/2}$, alpha unchanged.}
\begin{ruledtabular}
\begin{tabular}{d d d d}
\multicolumn{1}{c}{$A~(\mu\Phi_0 \, \text{Hz}^{-1/2})$} &
\multicolumn{1}{c}{$\alpha$} &
\multicolumn{1}{c}{$\tau_{\phi,R}$~(ns)} &
\multicolumn{1}{c}{$\tau_{\phi,E}$~(ns)} \\
\hline
 1.78 & 0.61 &  1.2 &   2.5 \\
 1.98 & 0.79 &  5.6 &  15.4 \\
 3.35 & 0.95 &  8.9 &  37.8 \\
\hline
0.178 & 0.61 & 20.9 &  43.2 \\
0.198 & 0.79 & 73.3 & 202.5 \\
0.335 & 0.95 & 99.5 & 400.8 \\
\hline
1.0   & 0.61 &  2.4 &   5.1 \\
1.0   & 0.79 & 11.9 &  33.1 \\
1.0   & 0.95 & 31.6 & 130.5 \\
\end{tabular}
\end{ruledtabular}
\end{table}

\section{Concluding remarks}

In conclusion, we have presented data showing that, in general, flux noise scales as $1/f^\alpha$, where $0.6 \lesssim \alpha \lesssim 1.0$. Our subsequent calculations show that the predicted dephasing times $\tau_{\phi,R}$ and $\tau_{\phi,E}$ of a qubit are very sensitive to the value of $\alpha$. As the value of $\alpha$ increases, both $\tau_{\phi,R}$ and $\tau_{\phi,E}$ increase dramatically---by an order of magnitude in some cases. Since experimentally inferred values of $S_\Phi(1~\text{Hz})$ from qubit measurements have generally assumed that $\alpha = 1$, a nonunity value of $\alpha$ can introduce a significant error into the inferred value of $A$. Furthermore, we have shown that while the lower cutoff frequency $f_1$ (set by the total measurement time) does not significantly affect $\tau_\phi$, the upper frequency cutoff $f_2$ can significantly change $\tau_\phi$ in a manner dependent on the value of $\alpha$, particularly for the echo sequence. Moreover, we have shown that by examining the directly measurable ratio $\tau_{\phi,E} / \tau_{\phi,R}$ and the dephasing function $g(t)$, experimentalists may have a probe into the values of $\alpha$ and $f_2$. Finally, the frequency at which the flux noise spectra pivot can dramatically affect the sensitivity of $\tau_\phi$ to $\alpha$.

Most importantly, these results demonstrate that lowering the flux noise amplitude is not the only method of increasing qubit dephasing times. With a more detailed understanding of what sets $\alpha$ experimentally---for, example, the geometry of the qubit washer---it may be possible to increase dephasing times substantially by raising the value of $\alpha$. Finally, we note that with straightforward modification our formalism could be used to calculate dephasing times from critical current noise and charge noise for the case $\alpha \neq 1$.

\section{Acknowledgements}

This research was funded by the CFN of the DFG and by the Office of the Director of National Intelligence (ODNI), Intelligence Advanced Research Projects Activity (IARPA), through the Army Research Office. All statements of fact, opinion or conclusions contained herein are those of the authors and should not be construed as representing the official views or policies of IARPA, the ODNI, or the U.S. Government.

\end{document}